\documentclass[
    aps,
    prd,
    twocolumn,
    superscriptaddress,
    nofootinbib
]{revtex4-2}

\usepackage{booktabs}
\usepackage{amsmath}
\usepackage{amssymb}
\usepackage{graphicx}
\usepackage{booktabs}
\usepackage{siunitx}
\usepackage{xcolor}
\usepackage{hyperref}

\usepackage{amsmath} 
\usepackage{orcidlink}
\usepackage{xcolor}
\usepackage[normalem]{ulem}

\usepackage{color}

\usepackage{savesym}
\savesymbol{tablenum}
\usepackage{siunitx}
\restoresymbol{SIX}{tablenum}
\usepackage{booktabs}

\begin{document}

\title{Field-level inference under pressure: the information content of the thermal Sunyaev--Zel'dovich field}

\author{Carmen~Embil Villagra\,\orcidlink{0009-0001-3987-7104}}
\affiliation{DAMTP, Centre for Mathematical Sciences, University of Cambridge, Wilberforce Road, Cambridge CB3 0WA, UK}
\affiliation{Kavli Institute for Cosmology Cambridge, Madingley Road, Cambridge CB3 0HA, UK}

\author{Adrian E. Bayer\,\orcidlink{0000-0002-3568-3900}}
\affiliation{The NSF AI Institute for Artificial Intelligence and Fundamental Interactions, Cambridge, MA 02139, USA}
\affiliation{Laboratory for Nuclear Science, Massachusetts Institute of Technology, Cambridge, MA 02139, USA}
\affiliation{Center for Astrophysics $|$ Harvard \& Smithsonian, 60 Garden Street, Cambridge, MA 02138, USA}
\affiliation{Perimeter Institute for Theoretical Physics, 31 Caroline Street North, Waterloo, Ontario N2L 2Y5, Canada}

\author{William~R.~Coulton}
\affiliation{Department of Physics, University of Oxford, Denys Wilkinson Building, Keble Road, Oxford OX1 3RH, UK.}

\author{Alina~Sabyr\,\orcidlink{0000-0003-3595-4384}}
\affiliation{Berkeley Center for Cosmological Physics, Department of Physics, University of California, Berkeley, CA 94720, USA}
\affiliation{Lawrence Berkeley National Laboratory, One Cyclotron Road, Berkeley, CA 94720, USA}

\author{Francisco Villaescusa-Navarro}
\affiliation{Center for Computational Astrophysics, Flatiron Institute, 162 5th Avenue, New York, NY, 10010, USA}
\affiliation{Department of Astrophysical Sciences, Princeton University, Princeton, NJ 08544, USA}

\author{Blake~D.~Sherwin}
\affiliation{DAMTP, Centre for Mathematical Sciences, University of Cambridge, Wilberforce Road, Cambridge CB3 0WA, UK}
\affiliation{Kavli Institute for Cosmology Cambridge, Madingley Road, Cambridge CB3 0HA, UK}

\author{Anthony~Challinor\,\orcidlink{0000-0003-3479-7823}}
\affiliation{Institute of Astronomy, Madingley Road, Cambridge CB3 0HA, UK}
\affiliation{DAMTP, Centre for Mathematical Sciences, University of Cambridge, Wilberforce Road, Cambridge CB3 0WA, UK}
\affiliation{Kavli Institute for Cosmology Cambridge, Madingley Road, Cambridge CB3 0HA, UK}

\author{J.~Colin~Hill}
\affiliation{Department of Physics, Columbia University, New York, NY 10027, USA}

\author{Nicholas~Battaglia}
\affiliation{Department of Astronomy, Cornell University, Ithaca, NY 14853, USA}

\begin{abstract}

The thermal Sunyaev--Zel'dovich (tSZ) field is highly non-Gaussian and contains cosmological information that cannot be fully captured by its angular power spectrum. In this work, we investigate the information content of this field using a convolutional neural network (CNN), and compare to simple and interpretable summary statistics, in a simulation-based inference (SBI) framework. We perform the first field-level analysis of the tSZ field, using a suite of $10{,}000$ simulated Compton-$y$ maps, where we jointly vary the cosmological parameters $\Omega_c$ and $\sigma_8$ (the density fraction for cold dark matter and the matter clustering amplitude, respectively) and two parameters describing the amplitude and shape of the electron pressure profile, $P_0$ and $\beta$, that aim to capture astrophysical uncertainties. We consider both noiseless simulations and simulations including realistic Simons Observatory (SO) noise after component separation. We use SBI to compare the constraints obtained from the CNN with those from the three Minkowski functionals (MFs), individually and in combination, as well as the angular power spectrum. This is the first application of SBI to the MFs of the tSZ field. We find that the CNN provides tighter constraints than conventional summary statistics in the noiseless case. However, at SO-like noise levels, the combination of MFs and the power spectrum recovers constraints comparable to those obtained with the CNN. Our results highlight the potential of these interpretable summary statistics, as well as field-level inference, for extracting information from the tSZ field, motivating efforts toward their application to real data.

\end{abstract}

\maketitle


\section{Introduction} \label{sec:intro}

The thermal Sunyaev--Zel'dovich (tSZ) effect \cite{1969Ap&SS...4..301Z,1970SZ} is a powerful probe of the late-time growth of cosmic structure through its sensitivity to the most massive objects in the Universe. It arises when cosmic microwave background (CMB) photons inverse-Compton scatter off hot, free electrons, primarily found in galaxy groups and clusters, producing a characteristic spectral distortion in the CMB. The number of these structures as a function of mass and redshift depends directly on cosmology through the halo mass function, making the tSZ effect particularly sensitive to the amplitude of matter fluctuations, commonly parametrized by $\sigma_8$ \cite{Bhattacharya_2012,PhysRevD.87.023527}. The tSZ signal is also sensitive to other cosmological parameters, with applications including constraints on the sum of neutrino masses, primordial non-Gaussianity, and the dark energy equation of state (e.g., \cite{PhysRevD.88.063526,10.1093/mnras/sty823,10.1093/mnras/staa1835,2026arXiv260706692A}).

However, extracting this cosmological information is challenging. The tSZ signal is dominated by galaxy groups and clusters, therefore the resulting field is highly non-Gaussian and much of its information cannot be captured by conventional two-point statistics alone \cite{Alina}. One approach is to use the imprint of the tSZ effect on our CMB maps to identify galaxy clusters and, using additional information about their masses and redshifts, extract constraints from the halo mass function directly \cite{Haiman_2001,2011ARA&A..49..409A,Hasselfield_2013,refId0,10.1093/mnras/stz2153,PhysRevD.110.083510}. 
These cluster number-count analyses have provided powerful cosmological constraints, but they are only sensitive to the most massive clusters, and therefore do not exhaust the full cosmological information present in the tSZ field. An alternative is to study the statistical properties of the tSZ field itself, without first identifying individual objects. 

The tSZ field can be obtained by combining CMB observations at different frequencies, using its distinct spectral distortion, for example through a needlet internal linear combination (NILC) \cite{2009A&A...493..835D}. A variety of summary statistics has been explored to extract information from component-separated tSZ maps, as well as modeling the tSZ signal present in individual frequency maps, for example: through the one-point probability distribution (PDF) \cite{hill2014atacamacosmologytelescopemeasurement,refId0_tsz,PhysRevD.99.103511}; power spectrum \cite{10.1046/j.1365-8711.2002.05889.x,PhysRevD.88.063526,10.1093/mnras/stx766,refId0_HL,refId0_SD,10.1093/mnras/sty823,refId0_tsz}; bispectrum \cite{Bhattacharya_2012,refId0_tsz,Coulton_2018}; skew-spectra \cite{10.1111/j.1365-2966.2011.19679.x,10.1093/mnras/sts232}; and moments \cite{PhysRevD.86.122005,PhysRevD.87.023527}. More recently, Ref.~\cite{Alina} compared the constraining power of several non-Gaussian summary statistics, including Minkowski functionals (MFs), peak counts, minima counts, and moments, finding the MFs to provide the strongest constraints among them. In this paper we aim to find the total information content of the tSZ field and estimate how much of this information can be recovered by these summary statistics.

One way to address this question is to compare the constraining power of these summary statistics with that of a neural network trained directly on tSZ simulated maps. Convolutional neural networks (CNNs) learn the complex, nonlinear features from simulated fields, combining information from a range of different scales that may be difficult to capture with conventional statistics. 
Early applications used CNN outputs for weak-lensing cosmology \cite{Fluri2018,Ribli2019,Fluri2019}. Subsequent work combined neural networks with simulation-based inference and information-theoretic training objectives \cite{Jeffrey2021,Sharma2024,Lanzieri2024}. Related approaches have also been applied to galaxy distributions \cite{2023mla..confE..18L,2022ApJ...937..115V, Bayer:2026zcr}, dark-matter density fields \cite{2017arXiv171102033R,2018arXiv180804728M,2025ApJ...989...47L}, 21-cm maps \cite{2025JCAP...11..082C,2019MNRAS.484..282G}, and the Lyman-$\alpha$ forest \cite{Nayak2024}. Here, we apply field-level inference to simulated tSZ maps to extract the full cosmological information from the field, which, to our knowledge, is the first time this has been done.

Applying such field-level inference directly to data requires simulations that accurately capture the cosmological and astrophysical processes, as well as realistic instrumental noise, and Galactic and extragalactic contaminants. Producing simulations with this level of realism is still challenging, particularly for observables like the tSZ effect whose properties depend strongly on the astrophysics that govern the intracluster gas distribution, and whose observations are affected by contamination from astrophysical foregrounds like the cosmic infrared background (CIB) and thermal dust emission from the Galaxy. For this reason, our aim is not to establish a CNN for direct application to data, but rather to benchmark the information content of the tSZ field. By comparing its constraining power to simpler, interpretable summary statistics, we can assess how much additional information the CNN is able to extract and whether that information can already be captured by more conventional approaches.

We compare field-level inference with the constraints obtained from the three MFs, individually and in combination, as well as the angular power spectrum. To do so, we use simulation-based inference (SBI), which allows us to infer posterior distributions directly from simulations, without the need to write down a likelihood (see \cite{doi:10.1073/pnas.1912789117} for a review, and e.g. \cite{10.1093/mnras/stab2215,PhysRevD.110.063510,Baxter_2024,drn7-ggqk,2025arXiv251216869D} for previous applications to CMB science). This is particularly important for non-Gaussian summary statistics like the MFs, for which the likelihood is not known analytically. Previous tSZ analyses of MFs and related statistics, such as the PDF, have relied on Gaussian approximations \cite{Alina,hill2014atacamacosmologytelescopemeasurement}; while Ref.~\cite{PhysRevD.99.103511} developed an accurate analytic model for the tSZ one-point PDF, including halo overlaps and clustering. More recently, SBI has been applied to investigate the validity of the Gaussian likelihood approximation for the tSZ power spectrum \cite{Xu2026}. Here, we apply SBI to the MFs, the power spectrum, and the summary representations extracted from the maps by the CNN, using the same suite of simulations for all three approaches.

Our analysis uses a suite of $10,000$ halo-based simulated tSZ maps varying the cosmological parameters $\sigma_8$ and $\Omega_c$ (the density fraction in cold dark matter). We note that the tSZ signal is also sensitive to the Hubble parameter $h$, however its dependence is much weaker than that on $\sigma_8$ and $\Omega_c$ (in particular, for the power spectrum: $C_\ell^{yy}\propto\sigma_8^{8.1}\Omega_m^{3.2}h^{-1.7}$ \cite{10.1093/mnras/sty823}). Therefore, we do not aim to constrain $h$, for which other cosmological probes are better suited. Crucially, we allow the parameters governing the amplitude and shape of the cluster pressure profiles, $P_0$ and $\beta$, respectively, to vary alongside the cosmological parameters. This provides an approximate way of accounting for uncertainties in the astrophysical modeling of the tSZ signal. We perform this analysis on both noiseless maps and maps including realistic noise levels expected for the Simons Observatory (SO) after component separation. 

The paper is organized as follows. In Sec.~\ref{sec:sims}, we describe the simulated tSZ maps and the treatment of SO noise. In Sec.~\ref{sec:summary-stats}, we introduce the summary statistics considered in this work and the CNN architecture used for field-level inference. In Sec.~\ref{sec:sbi}, we describe their common training and posterior inference framework. We present our results in Sec.~\ref{sec:results}, comparing the cosmological and astrophysical constraints obtained from the different methods, both with and without noise. Finally, in Sec.~\ref{sec:discussion}, we discuss our findings, their limitations, and directions for future work. Appendix~\ref{app:results-cosmo} presents idealized results for which the astrophysical parameters are assumed to be known perfectly, and Appendix~\ref{app:coverage-tests} describes the coverage tests we used to assess the reliability of the inferred parameter distributions.

Throughout this paper, unless stated otherwise, we assume a spatially flat $\Lambda$CDM cosmology based on constraints from \emph{Planck} \cite{Planck_cosmo_2018}, with fiducial parameters set to the following: present-day baryon density $\Omega_{b}=0.0493$ and cold dark matter density $\Omega_c=0.264$; $\sigma_{8}=0.811$; primordial power spectrum spectral index $n_{s}=0.965$, optical depth to reionization $\tau_{\rm reio}=0.054$, and Hubble constant $H_{0}=67.4\,\text{km}\,\text{s}^{-1}\,\text{Mpc}^{-1}$. We assume no massive neutrinos, $\sum m_{\nu}=0\,\text{eV}$.

\section{Simulations}\label{sec:sims}

The tSZ effect produces a characteristic distortion of the CMB temperature, which, in the non-relativistic limit, can be expressed at a given frequency $\nu$ and sky position $\hat{\boldsymbol{n}}$ as
\begin{equation}
    \frac{\Delta T(\nu,\hat{\boldsymbol{n}})}{T_{\text{CMB}}} = g(\nu) y(\hat{\boldsymbol{n}}) \; ,
\end{equation}
where $T_{\text{CMB}}$ is the CMB temperature today. The dependence on the photon frequency is given by the tSZ spectral function $g(\nu) =  x \coth(x/2) -4$, where $x = (h\nu)/(k_B T_{\text{CMB}})$, $k_B$ is the Boltzmann's constant, and  $h$ is the Planck's constant. The amplitude of the signal is determined by the dimensionless Compton-$y$ parameter, which is proportional to the electron pressure integrated along the line of sight:
\begin{equation}
    y(\hat{\boldsymbol{n}}) = \frac{\sigma_{\text{T}}}{m_e c^2} \int P_e(\hat{\boldsymbol{n}},l) dl \, ,
\end{equation}
where $P_e$ is the electron pressure, $m_e$ is the electron mass, $\sigma_{\text{T}}$ is the Thomson cross section, and $c$ is the speed of light. Throughout this work, we perform our analysis on simulated Compton-$y$ maps, which we describe in this section.

We generate these $y$ maps using \verb|hmpdf|\footnote{\url{https://github.com/leanderthiele/hmpdf}}\cite{PhysRevD.99.103511, Thiele2020}. Detailed descriptions of the simulations and numerical convergence tests for different summary statistics can be found in Ref. \cite{Alina} and Ref. \cite{PhysRevD.99.103511}. Here, we provide a brief overview and list the numerical settings used in producing the maps. We note that all cosmological quantities in \verb|hmpdf| are computed using \verb|CLASS|~\cite{CLASS_CODE}.

\verb|hmpdf| Poisson samples halos from the Tinker et al.\ halo mass function~\cite{Tinker2010} with the following integration limits: $M_{\rm 200m}=10^{11}$--$10^{16}M_{\odot}$ and $z=0.005$--$6$, where $M_{\rm 200m}$ corresponds to the spherical overdensity mass defined as the mass within radius $R_{200\rm m}$, where the density is 200 times the mean matter density at a given redshift, $z$. The mass function is discretized in $N_{z}=150$ redshift and $N_{M}=150$ mass bins using Gauss--Legendre quadrature, with mass bins defined logarithmically. The sampled halos are distributed randomly across the simulated sky area.

Halos in a given redshift and mass bin are assigned a tSZ profile using the electron pressure profile from Ref.~\cite{Battaglia2012}, which has a generalized Navarro-Frenk-White (GNFW) form:
\begin{equation}
    P_{e}(x, M_{\rm 200c},z)=\frac{P_{\rm 200c}(z)}{1.932}\frac{P_{0}(x/x_{\rm c})^{\gamma}}{(1+(x/x_{\rm c})^{\alpha})^{\beta}}\, ,
\end{equation}
where, here, $x=r/R_{\rm 200c}$ and $M_{\rm 200c}$ denotes the spherical overdensity mass within radius $R_{\rm 200c}$, where the density is 200 times the critical density $\rho_{\rm c}(z)$. The parameter $P_{0}$ is the profile amplitude, $x_{\rm c}$ is the core scale, and $\beta$ is the outer slope with the following mass and redshift dependencies:
\begin{align}
    P_{0}(M_{\rm 200c},z)&=P^{\rm amp}_{0}\left(\frac{M_{\rm 200c}}{10^{14}M_{\odot}}\right)^{0.154}(1+z)^{-0.758} \, , \nonumber \\
    x_{\rm c}(M_{\rm 200c},z)&=0.497\left(\frac{M_{\rm 200c}}{10^{14}M_{\odot}}\right)^{-0.00865}(1+z)^{0.731} \, , \\
    \beta(M_{\rm 200c},z)&=\beta^{\rm amp}\left(\frac{M_{\rm 200c}}{10^{14}M_{\odot}}\right)^{0.0393}(1+z)^{0.415} \, , \nonumber
\end{align}
with fiducial values $P_{0, \rm fid}^{\rm amp}=18.1$ and $\beta_{\rm fid}^{\rm amp}=4.35$. The other shape parameters $\gamma=0.3$ and $\alpha=1$ are fixed. The quantity $P_{\rm 200c}(z)$ is the self-similar pressure amplitude:
\begin{equation}
    P_{\rm 200c}(z)=\frac{200GM_{\rm 200c}\rho_{\rm c}(z)\Omega_{b}}{2R_{200c}\Omega_{m}} \, ,
\end{equation}
where $G$ is the gravitational constant and $\Omega_m$ is the present-day matter density parameter, $\Omega_m = \Omega_b + \Omega_c$. The profiles are fit to the thermal gas pressure in hydrodynamical simulations. The thermal gas and electron pressure are related via $P_{\rm th}(x)\approx1.932P_{e}(x)$ (assuming fully ionized gas). To convert between $M_{200\rm m}$ in the mass function and $M_{200\rm c}$ in the pressure profiles, we use the halo concentration--mass relation from
Ref.~\cite{Duffy2008}. Finally, we truncate the pressure profiles at $R_{\rm max}=2R_{\rm vir}$, where $R_{\rm vir}$ is the virial radius, which encloses a mean density of  $\Delta_{\rm vir}(z)\rho_{\rm c}(z)$, as opposed to $200\rho_{\rm c}(z)$ for $R_{200c}$. $\Delta_{\rm vir}(z)$ is the redshift-dependent virial overdensity \cite{Bryan_1998}.

Hereafter, we use $P_0$ and $\beta$ to denote the amplitude parameters $P_0^{\rm amp}$ and $\beta^{\rm amp}$, respectively; their prescribed mass and redshift dependencies remain fixed.

We simulate maps with pixel resolution of 0.5\,arcmin and area $f_{\rm sky}=0.003$ (i.e., each map has an area of approximately $11 \times 11$ deg$^{2}$). We choose this sky area to keep the computation of the summary statistics and CNN training manageable, both in terms of computational time and memory, while still being able to capture sample variance with a reasonable number of simulations, which has been shown in Ref.~\cite{Alina}. To capture the astrophysical uncertainty in the modeling of the amplitude and shape of the electron pressure profiles, we choose to vary $P_{0}$ and $\beta$ alone based on the results in Ref.~\cite{Sabyr2027}, which have shown that, in the noiseless case, the total reduction in cosmological constraining power from additionally varying the $x_{c}$ amplitude is comparable to the case of varying the amplitude of $P_{0}$ and $\beta$ alone. 

To create the suite of simulations used in the training of our compression and inference frameworks (described in Sec.~\ref{sec:summary-stats} and Sec.~\ref{sec:sbi}, respectively), we sample $P_0$ and $\beta$ together with the cosmological parameters $\Omega_c$ and $\sigma_8$ from Sobol sequences \cite{SOBOL1967}. To do this efficiently, we generate 48 Sobol sequences using different random seeds, each containing $2^8$ points in the four-dimensional parameter space. We sample from the following bounds: $\Omega_{c}=[0.1,0.4]$, $\sigma_{8}=[0.6, 1.0]$, $P_{0}=[0.8\times P_{0, \rm fid}^{\rm amp}, 1.2\times P_{0, \rm fid}^{\rm amp}]$, and $\beta=[0.8\times \beta_{\rm fid}^{\rm amp}, 1.2\times\beta_{\rm fid}^{\rm amp}]$. This sampling corresponds to a joint uniform prior over these ranges, with no imposed correlations between the parameters. The 20\% variation we adopt around the fiducial values of the pressure-profile parameters is motivated by observational constraints on the properties of massive galaxy clusters (e.g. \cite{10.1093/mnras/stz2153,2010A&A...517A..92A,He_2021,PhysRevD.105.123525}), which dominate the tSZ signal. However, broader priors may be needed to capture the larger astrophysical uncertainties for the lower-mass halos in our simulations. We leave this investigation for future work.

In Appendix~\ref{app:results-cosmo}, we show the analysis performed on simulations that vary only cosmology and keep the astrophysical parameters fixed. In that case, we use the following bounds based on the analysis in Ref.~\cite{Alina}: $\Omega_{c}=[0.2,0.32]$ and $\sigma_{8}=[0.76, 0.86]$.

In the `noiseless' case we smooth the maps with a Gaussian beam with $\theta_{\rm FWHM}=1.4$ arcmin. In the `noisy' case, we additionally combine the $y$ maps with an independent realization of Gaussian noise for each map that is consistent with levels expected from post-component separation for SO (i.e., residual foregrounds plus instrumental noise). We follow the same procedure as in Ref.~\cite{Alina} and use the tSZ noise predicted using the standard internal linear combination (ILC) component separation method assuming $f_{\rm sky}=0.4$ \cite{SOnoise2019}\footnote{\url{https://github.com/simonsobs/so_noise_models/tree/master/LAT_comp_sep_noise/v3.1.0}}\footnote{Note that these noise levels are appropriate for the nominal SO Large Aperture Telescope survey, not the enhanced survey forecast in Ref.~\cite{SimonsObservatory:2025wwn}.}. To avoid extrapolating noise power spectra at high and low $\ell$, we filter both the signal and noise at $\ell\leq80$ and $\ell\geq7950$. Finally, we also Wiener filter the noisy tSZ maps with the filter $F_{\ell}^{\rm Wiener}=C_{\ell}^{\rm signal}/(C_{\ell}^{\rm signal}+C^{\rm noise}_{\ell})$, where $C_\ell^{\text{signal}}$ and $C_\ell^{\text{noise}}$ are the angular power spectra of the tSZ signal and the noise, respectively.

\section{Summary statistics and CNN} \label{sec:summary-stats}

Motivated by the findings in Ref.~\cite{Alina}, we limit our summary statistics to MFs and the angular power spectrum. In this section, we present the ResNet architecture used for the CNN (see Sec.~\ref{subsec:cnn-summary}), as well as the computation of the MFs (see Sec.~\ref{subsec:mfs}) and the angular power spectrum (see Sec.~\ref{subsec:power-spectrum}).

In each of the three cases, we obtain a summary representation of the input map: while the CNN learns this representation directly from the field, the MFs and power spectrum are predefined summary vectors. Throughout this section, we also present how we further compress each of these summary representations by using a multilayer perceptron (MLP), before inferring the parameters of interest. The posterior inference framework common to all approaches is described in Sec.~\ref{sec:sbi}.

\subsection{Field-level CNN} \label{subsec:cnn-summary}

We adopt a ResNet-18 architecture \cite{He2016} as our map compressor, following its use for neural compression in weak-lensing inference \cite{Jeffrey2021,Lanzieri2024}. Reference~\cite{Sharma2024} compared ResNet-18, ResNet-34, and ResNet-50 and found similar constraining power in their weak-lensing analysis, motivating our use of the smaller ResNet-18 model.

We initialize the encoder with weights pretrained on ImageNet \cite{5206848,russakovsky2015imagenetlargescalevisual}. The first convolutional layer is modified to accept single-channel tSZ maps rather than three-channel RGB images, with its filters initialized by averaging the pretrained weights across the three color channels. We replace the final classification layer with an MLP consisting of a fully connected layer with 256 hidden units, a ReLU activation, dropout, and a linear layer producing a 128-dimensional representation. The resulting encoder maps each input field $\mathbf{z}$ to a learned summary $\mathbf{s}=f_\psi(\mathbf{z})$, where $\psi$ comprises the trainable parameters of both the ResNet and its final MLP. The encoder is trained jointly with the posterior estimator, as described in Sec.~\ref{sec:sbi}.

\begin{figure*}[t]
    \centering 
    \includegraphics[width=\textwidth]{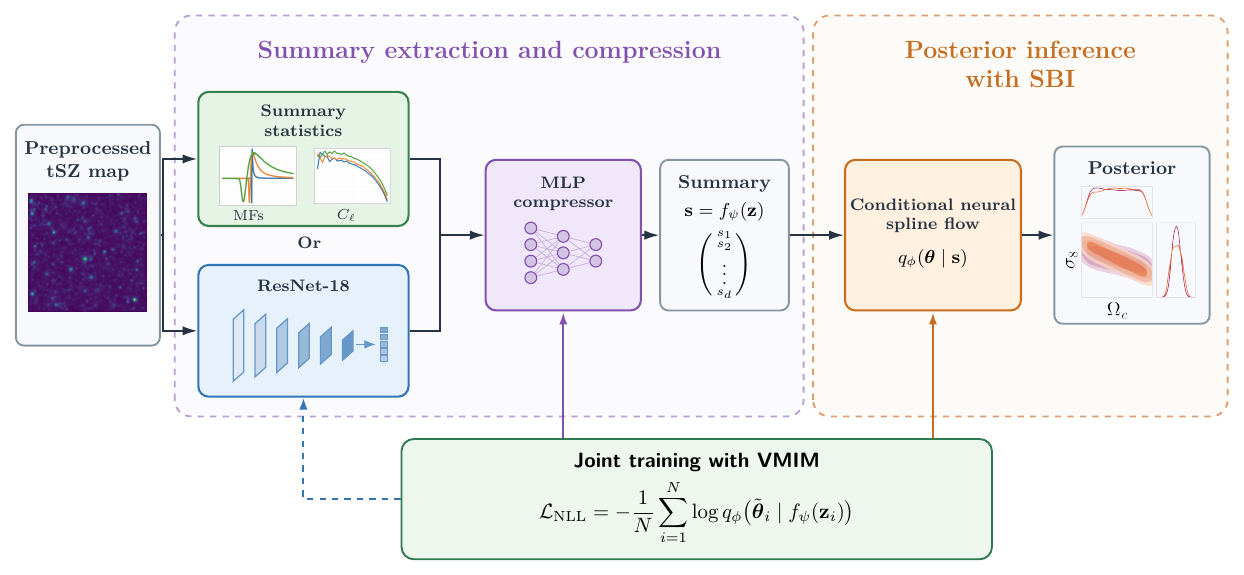}
    \caption{Schematic representation of the inference framework used in this work. For each simulated tSZ map, we extract a summary representation either directly from the field using a ResNet-18 or through predefined summary statistics. In both cases, the resulting representation is compressed using an MLP, producing a low-dimensional summary. The posterior is then inferred with SBI, by using this summary to condition an NSF. The trainable components are optimized jointly using VMIM. In the ResNet-18 case, this also includes its trainable parameters.
}
    \label{fig:cnn_architecture}
\end{figure*}

\subsection{Minkowski functionals} \label{subsec:mfs}

Minkowski functionals are morphological descriptors of a field that characterize the abundance, geometry, and topology of regions above a given threshold \cite{Hadwiger1957,MFs_Hermann,Schmalzing1996,Mecke1994,SchmalzingGorski1998}. We express the threshold in normalized units, defining $\nu = y_{\rm th}/\sigma_{\rm fixed}$, where $y_{\rm th}$ is the threshold in Compton-$y$ units and $\sigma_{\rm fixed} = 1.5 \times 10^{-6}$ is the average root mean squared of the 1000 tSZ maps closest to the fiducial cosmology\footnote{We note that changes in the number of maps used to compute $\sigma_{\rm fixed}$ do not impact our results.}. For a two-dimensional field such as the tSZ Compton-\(y\) map, there are three independent Minkowski functionals: \(V_0\), \(V_1\), and \(V_2\). These describe, respectively, the area fraction above the threshold, 
\begin{equation}
    V_0(\nu) = \frac{1}{A} \int_{\Sigma(\nu)}da \; ,
\end{equation}
where $A$ is the total area of the map, $da$ is the area element and $\Sigma(\nu) = \{y/\sigma_{\text{fixed}}>\nu \}$ (so $V_0$ is effectively one minus the cumulative distribution function); the total boundary length of these regions,
\begin{equation}
    V_1(\nu) = \frac{1}{4A} \int_{\delta\Sigma(\nu)}dl \; ,
\end{equation}
where $dl$ is the boundary length element and $\delta \Sigma(\nu)$ is the excursion set boundary; and their topology, characterized by the genus or Euler characteristic,
\begin{equation}
    V_2(\nu) = \frac{1}{2 \pi A} \int_{\delta\Sigma(\nu)} \mathcal{K} dl \; ,
\end{equation}
where $\mathcal{K}$ is the curvature of the boundary. By measuring these quantities over a range of thresholds, we can characterize the global morphological properties of the field. 

We compute the three MFs using the \texttt{LensTools} package \cite{Petri2016}\footnote{\url{https://github.com/apetri/LensTools}}. In practice, the three functionals are evaluated directly from the
field and its spatial derivatives as
\begin{equation}
\begin{aligned}
V_0(\nu) &=
\frac{1}{A}
\int_A
\Theta\bigl(y(\mathbf{p})-\nu\bigr)
\,d^2\mathbf{p} \; ,
\\[6pt]
V_1(\nu) &=
\frac{1}{4A}
\int_A
\delta_{\rm D}\bigl(y(\mathbf{p})-\nu\bigr)
\sqrt{y_i^2+y_j^2}
\,d^2\mathbf{p} \; ,
\\[6pt]
V_2(\nu) &=
\frac{1}{2\pi A}
\int_A
\delta_{\rm D}\bigl(y(\mathbf{p})-\nu\bigr)
\\
&\quad\times
\left(\frac{
2y_i y_j y_{ij}
-y_i^2 y_{jj}
-y_j^2 y_{ii}
}{
y_i^2+y_j^2
}\right)
\,d^2\mathbf{p} \; ,
\end{aligned}
\label{eq:minkowski_functionals}
\end{equation}
where $A$ is the map area, $\mathbf{p}=(i,j)$ denotes the position on the map, $\Theta$ is the Heaviside step function, and $\delta_{\mathrm{D}}$ is the Dirac delta function. The quantities $y_i$ and $y_j$ denote the first spatial derivatives of the field along the two orthogonal coordinate directions, while $y_{ii}$, $y_{jj}$, and $y_{ij}$ denote its second derivatives. These expressions are evaluated numerically on the pixelized maps using centered finite differences to calculate the first and second spatial derivatives of the field. The Dirac delta functions entering \(V_1\) and \(V_2\) are approximated using finite-width threshold bins, with the corresponding integrands averaged over the pixels within each bin and divided  its width.

For each map, we evaluate the Minkowski functionals at 100 thresholds over the interval $-5 \leq \nu \leq 7$ . The thresholds are distributed using a hyperbolic sine transformation with a stretch parameter of 3, providing a finer sampling around \(\nu=0\), where the functionals vary most rapidly. Each functional is therefore represented by a vector of 100 values, giving a combined 300-dimensional summary when all three MFs are used jointly in the inference.

For the Minkowski functionals, we consider each of $V_0$, $V_1$, and $V_2$ individually, as well as their combination. When training on a single functional, the 100-dimensional input vector is compressed through an MLP with hidden layers of 64 and 32 neurons, producing a 20-dimensional representation. When combining all three functionals, their 300 values are concatenated and compressed using a larger MLP with hidden layers of 256 and 128 neurons, producing a 60-dimensional representation. Both architectures use GELU activations, and a dropout rate of 0.05. These MLP parameters are learned jointly with the posterior estimator as described in Sec.~\ref{sec:sbi}.

\subsection{Angular power spectrum}\label{subsec:power-spectrum}

In the flat-sky approximation, the power spectrum is defined through
\begin{equation}
\langle \tilde y^*(\boldsymbol{\ell})\tilde y(\boldsymbol{\ell}')\rangle
= (2\pi)^2\delta^{(2)}_{\rm D}(\boldsymbol{\ell}-\boldsymbol{\ell}')C_\ell^{yy} \; , 
\end{equation}
where \(\tilde y\) denotes the Fourier transform of the Compton-\(y\) field and \(\delta^{(2)}_{\rm D}\) is the 2D Dirac delta function. We estimate the power spectrum using \texttt{LensTools}, averaging over linearly spaced multipole bins of width \(\Delta\ell=100\). Following Ref.~\cite{Alina}, we keep 77 bins spanning \(125\leq\ell\leq7825\), with the first and last bins centered at \(\ell=175\) and \(\ell=7775\), respectively.

As for the MFs, we use an MLP to compress the measured power spectrum into a lower-dimensional representation. We pass the 77-dimensional vector through an MLP with two hidden layers of 64 and 32 neurons, GELU activations, and a dropout rate of 0.05, producing a 20-dimensional learned summary. In the case of the combination of all MFs and the power spectrum, we use the same MLP as the one described for the case of all MFs. Again, the MLP parameters are optimized jointly with the posterior estimation, following the training procedure described in Sec.~\ref{sec:sbi}.

\section{Posterior inference and training} \label{sec:sbi}

Extracting cosmological constraints from these summaries requires modeling their dependence on the cosmological and astrophysical parameters. Conventional analyses often assume a multivariate Gaussian likelihood, as in Ref.~\cite{Alina}. However, this approximation is not necessarily accurate for statistics such as MFs\footnote{The impact of assuming a Gaussian likelihood for the tSZ power spectrum has recently been investigated in \cite{Xu2026} (see also earlier work in Ref.~\cite{Zhang:2007psa}), finding that, although the likelihood is significantly non-Gaussian, this approximation has little impact on the inferred cosmological constraints for a Planck-like analysis restricted to multipoles $\ell<1000$.}. Rather than assuming a particular analytical form for the likelihood, we use SBI to learn the posterior distribution of the parameters directly from simulated maps and their summaries, allowing the same flexible inference framework to be applied to all methods.

We adopt the same variational mutual information maximization (VMIM) framework for all approaches. Previous studies have shown that VMIM can retain more cosmological information than conventional training objectives such as mean squared error (MSE), achieving performance comparable to more complex generative models \cite{Sharma2024,Lanzieri2024}. As introduced in Sec.~\ref{sec:summary-stats}, the main difference between the different approaches in our analysis lies in how the initial representation of the map is obtained: for field-level inference, the ResNet learns this representation directly from the map, whereas for the MFs and power spectrum, this is a predefined summary statistic. In all cases, this representation is further compressed using an MLP to obtain a lower-dimensional summary ($\mathbf{s}$), which is then used to condition a normalizing flow implemented using $\texttt{nflows}$\footnote{\url{https://github.com/bayesiains/nflows}}(see Sec.~\ref{subsec:normalizing-flow}). 

To describe all approaches within the same framework, we write $\mathbf{s}=f_\psi(\mathbf{z})$, where $\mathbf{z}$ denotes the input map for the CNN or the precomputed summary-statistic vector for the MFs and power spectrum. The trainable parameters $\psi$ therefore include both the ResNet and MLP for the CNN, but only the MLP for the MFs and power spectrum. A visual representation of the full architecture used in this work can be found in Fig.~\ref{fig:cnn_architecture}.

\subsection{Conditional normalizing flow}\label{subsec:normalizing-flow}

We use a neural spline flow (NSF) \cite{Durkan2019} to model the joint posterior distribution of the four parameters of interest, conditioned on the summary representations described in Sec.~\ref{sec:summary-stats}. The flow transforms a four-dimensional Gaussian distribution into a flexible, non-Gaussian posterior through a sequence of invertible transformations, allowing it to capture correlations between parameters. Specifically, we use five successive autoregressive spline transformations, with random permutations of the parameter ordering between them to capture dependencies across all dimensions. Each transformation uses eight spline bins and is parameterized by a neural network with two residual blocks and 128 hidden units.

The complete model therefore represents the posterior as
\begin{equation}
q_{\phi,\psi}(\boldsymbol{\theta}\mid\mathbf{z})
=
q_\phi(\boldsymbol{\theta}\mid f_\psi(\mathbf{z}))\; ,
\end{equation}
where $\psi$ and $\phi$ denote the trainable parameters of the compressor and the normalizing flow, respectively, and $\boldsymbol{\theta}$ the inferred parameters.

\subsection{VMIM objective and optimization}

The compressor and NSF are trained jointly by maximizing a variational lower bound on the mutual information between the parameters and the compressed representation \cite{BarberAgakov2003,Jeffrey2021,Lanzieri2024}. For a fixed training prior, this is implemented by minimizing the negative log-probability of the true simulation parameters under the learned conditional density:
\begin{equation}
\mathcal{L}_{\mathrm{NLL}}(\psi,\phi)
=
-\frac{1}{N}
\sum_{i=1}^{N}
\log q_\phi
\left(
\widetilde{\boldsymbol{\theta}}_i
\mid f_\psi(\mathbf{z}_i)
\right)\; ,
\end{equation}
where $\widetilde{\boldsymbol{\theta}}_i$ denotes the true parameters standardized using the training-set means and standard deviations. This objective forces the compressor to retain information about the joint parameter posterior, rather than only a point estimate.

Before training, the measured MFs summaries are standardized using the mean and standard deviation of each input bin, calculated exclusively from the training simulations. 
For the power spectrum, we apply a logarithmic transformation to the binned power spectrum and standardize each bin using the mean and standard deviation calculated as well only from the training simulations. The inferred parameters are also normalized using the mean and standard deviation of the training set. The simulations and their train--validation split are kept consistent across all methods.

For the CNN-based model, training is performed using the AdamW optimizer \cite{LoshchilovHutter2019}, with a batch size of 64 and gradient clipping at a maximum norm of 5. We use Optuna \cite{Akiba2019} to optimize the learning rate, weight decay, and dropout probability. The learning rate is reduced by a factor of two when the validation loss does not improve for twelve epochs. The best-performing models are selected using the best mean validation loss.

Once trained, each model can generate posterior samples by evaluating its learned low-dimensional representation and conditioning the normalizing flow on it. Posterior samples are then obtained directly, without requiring Markov chain Monte Carlo sampling.


\section{Results} \label{sec:results}

In this section, we compare the parameter constraints obtained from the CNN, MFs, and
power spectrum. We first consider noiseless maps in Sec.~\ref{sec:results-astro}, then include the SO post-component-separation noise described in Sec.~\ref{sec:results-noisy}. Results with fixed astrophysical parameters are presented in Appendix~\ref{app:results-cosmo}.

To assess the constraining power of each method, we evaluate the trained posterior distributions on 1,000 independent tSZ realizations generated at the same fiducial cosmological and astrophysical parameters: $\sigma_8 = 0.811$, $\Omega_c = 0.264$, $P_0 = 18.1$, and $\beta = 4.35$. For each realization, we draw 5,000 samples from the posterior inferred using each of the methods (CNN, power spectrum, and each of the MFs summaries). We then construct an equally weighted mixture of each of the posteriors across realizations, ensuring that each map contributes equally. We emphasize that we do not combine the information from the 1,000 maps into a joint posterior. Instead, this mixture allows us to compare the performance of the different methods at one fiducial set of parameters, averaging over both the posterior uncertainties and variations between maps. We note that these mixture distributions are then broader than the posteriors expected for individual realizations, as they incorporate both the uncertainty within each posterior and the scatter in their inferred means across realizations. However, this reduces our dependence on any particular realization, making it well suited to comparing the relative constraining power of the different methods. This also allows us to assess whether the resulting distributions are centered on the fiducial parameters.

Although our main goal is to compare the relative constraining power of the different methods, our results also provide an indication of the constraints potentially achievable with a larger survey. As discussed above, in the Gaussian limit, the mixture distributions are expected to be broader than typical individual posteriors by a factor of $\sqrt{2}$. Furthermore, a survey with SO-like sky coverage ($f_{\text{sky}}\sim 0.4$) could yield constraints roughly $\sqrt{0.4/0.003}\simeq 11.5$ times tighter than those from an individual map in our simulations. Combining both factors, the widths of the posterior distributions shown in our figures would be roughly $\sqrt{2}\times 11.5\simeq 16.3$ times smaller for an SO-like survey. This is, however, only an approximate estimate, particularly given some of our posteriors are limited by the prior boundaries and, in some cases, not Gaussian.

More tests on these inferred posteriors, including coverage tests as a function of number of training simulations, are shown in Appendix~\ref{app:coverage-tests}. The resulting distributions are estimated using Gaussian kernel density estimation (KDE), and the contours shown throughout this section enclose 68\% of the estimated probability density.

\subsection{Noiseless}\label{sec:results-astro}

\begin{figure*}
    \centering 
    \includegraphics[width=0.8\linewidth]{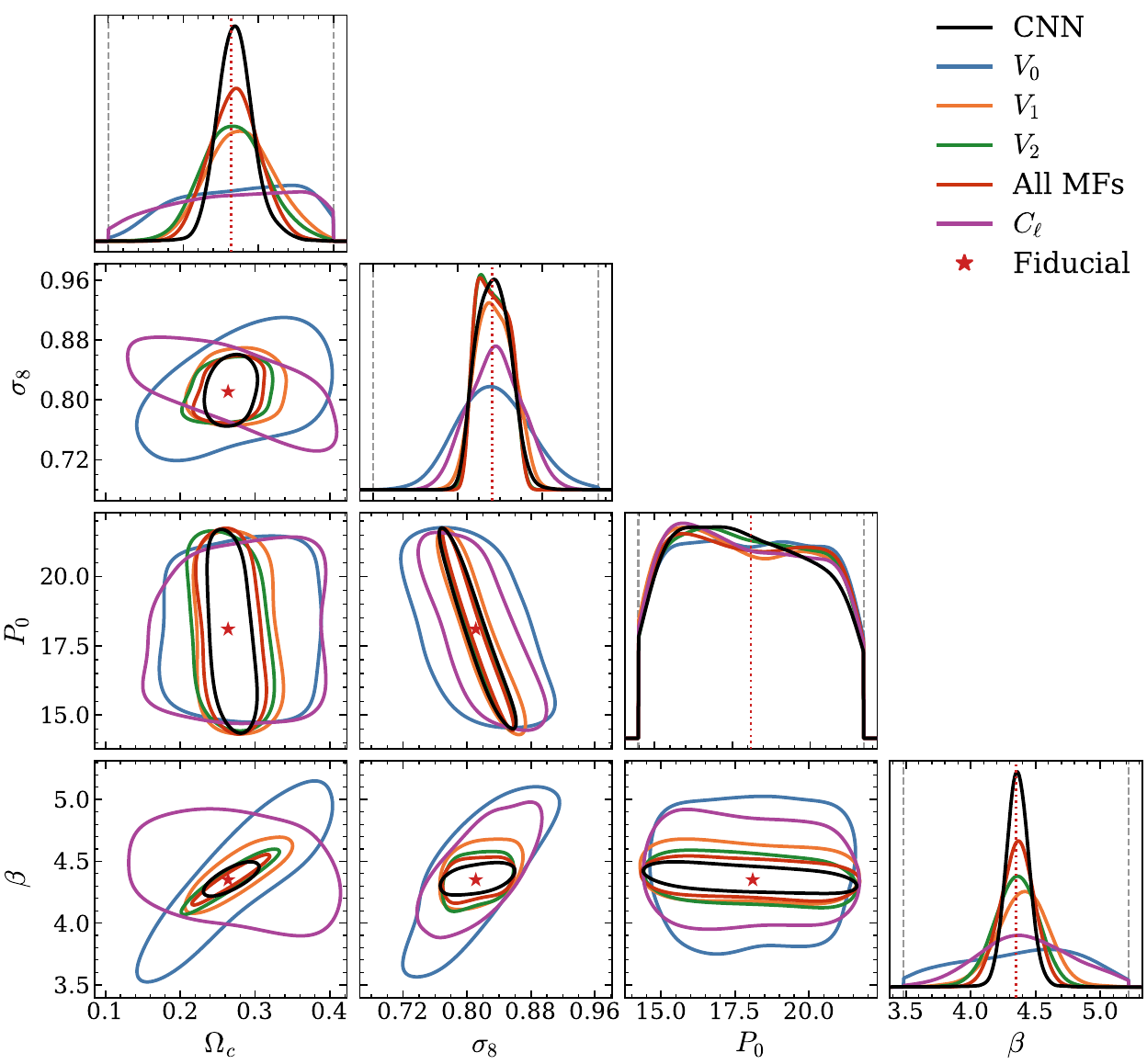}
    \caption{Marginalized posterior distributions for the cosmological parameters $\Omega_c$ and $\sigma_8$ and the astrophysical parameters $P_0$ and $\beta$, obtained from the CNN, the angular power spectrum, and the three Minkowski functionals (individually and combined) using \textbf{noiseless} simulations. The CNN generally provides the tightest constraints, particularly on $\Omega_c$ and $\beta$, with the combination of all three MFs yielding the next strongest constraints. The dashed regions denote the priors used on each parameter and the red $\star$ are the true parameter values.
    }
    \label{fig:results-noiseless}
\end{figure*}

We first consider the idealized case of noiseless maps described in Sec.~\ref{sec:sims}. The resulting constraints on all four varied parameters are shown in Fig.~\ref{fig:results-noiseless}. We compare the field-level CNN with the three MFs, individually and in combination, as well as with the angular power spectrum.

We find that the CNN generally provides the tightest constraints, particularly on $\Omega_c$ and $\beta$. Among the conventional summary statistics, the combination of all three MFs performs best, producing constraints that, although slightly broader than those obtained with the CNN, remain considerably tighter than those from the power spectrum. Individually, $V_1$ and $V_2$ already recover much of the constraining power of the combined MFs, while $V_0$ provides substantially weaker constraints.

We find that $P_0$ remains poorly constrained by all methods, with broad, nearly flat posteriors. As can be seen in the figure, due to the degeneracy between $P_0$ and $\sigma_8$, the prior on $P_0$ is directly influencing the total width of the $\sigma_8$ posterior.  We note, however, that observational constraints on the pressure profiles of massive galaxy clusters already exist, particularly from X-ray measurements \cite{2010A&A...517A..92A} and their combination with tSZ \cite{Planck2013_P0}. Furthermore, lensing measurements with galaxies or the CMB can provide independent estimates of cluster masses, allowing the calibration of the relation between the tSZ signal and halo mass, which can in turn help constrain the pressure-profile amplitude $P_0$. The complementarity between tSZ and CMB lensing measurements has been explored both through their angular cross-power spectrum \cite{Hill:2013dxa,PhysRevD.109.023529} and in cluster number-count analyses \cite{10.1093/mnras/stz2153}. 

Interestingly, while $\sigma_8$ exhibits a strong degeneracy with $P_0$, the constraints on $\Omega_c$ are only weakly affected by the uncertainty in this parameter, as shown by the nearly vertical contours in the $(\Omega_c,P_0)$ plane. Instead, $\Omega_c$ exhibits a much stronger degeneracy with $\beta$. This may be explained by the fact that a larger $\beta$ (sharper profile) can be compensated by a larger $\Omega_c$ and vice versa: a higher $\Omega_c$ value decreases the angular diameter distance, increasing the apparent angular size of halos, which can be compensated by a larger $\beta$.

Regarding the degeneracy directions of the different methods, we find that in most cases all summary statistics and the CNN are sensitive to similar degeneracies between parameters. There are two noticeable exceptions: the power spectrum exhibits opposite degeneracy directions in both the $(\Omega_c,\beta)$ and $(\Omega_c,\sigma_8)$ joint posteriors compared to MFs and the CNN. We exploit these different degeneracy directions in the next section, combining the MFs and power spectrum to improve their individual constraining power.

\subsection{With noise}\label{sec:results-noisy}

\begin{figure*}
    \centering 
    \includegraphics[width=0.8\linewidth]{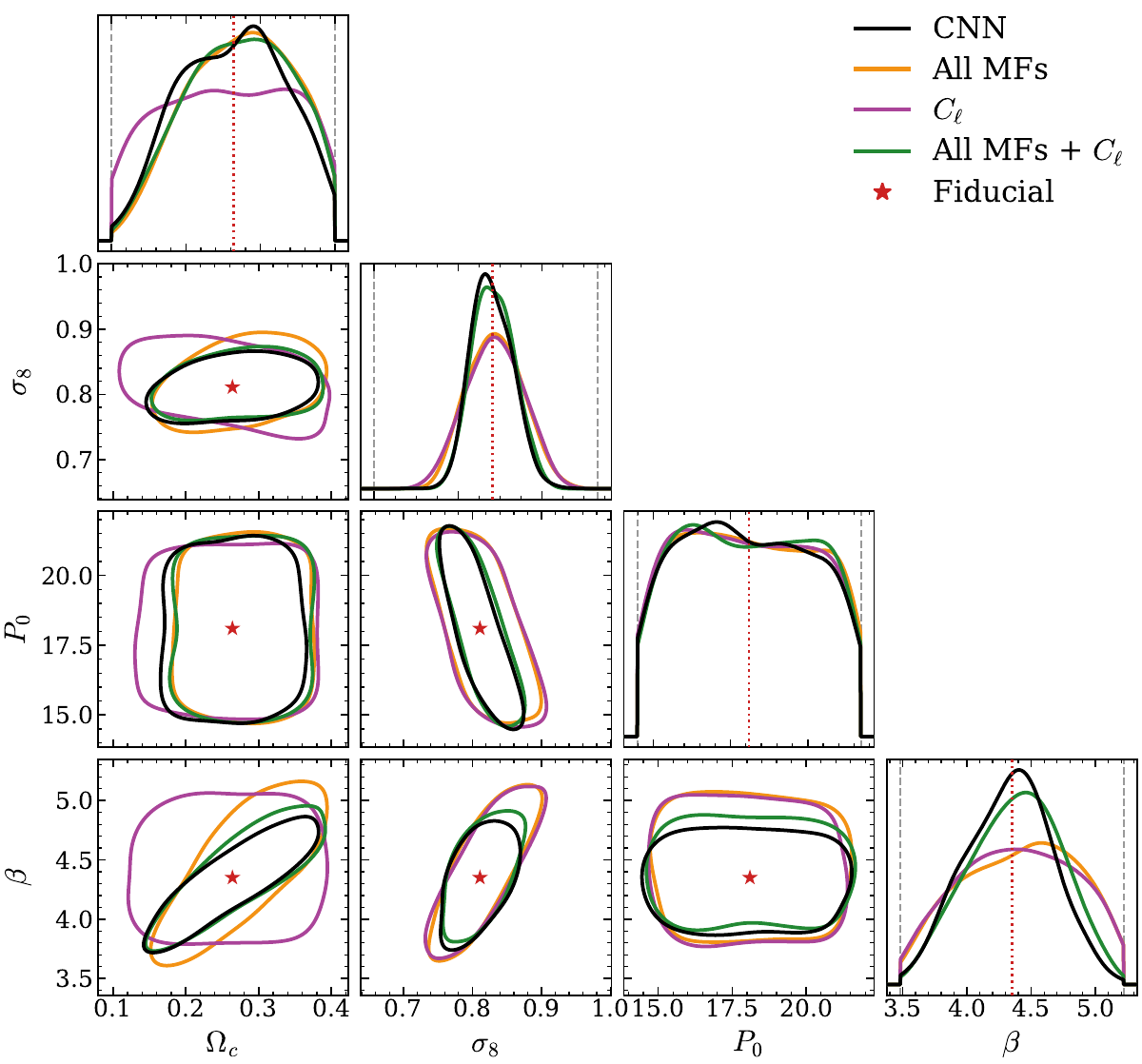}
    \caption{Same as Fig.~\ref{fig:results-noiseless} but with posteriors trained on simulations including realistic SO noise after component separation. The combination of all MFs and the power spectrum yields cosmological constraints comparable to those obtained with the CNN, recovering essentially all of its constraining power on cosmological parameters, while still performing slightly worse for the profile parameter $\beta$. We remind the reader that these constraints are obtained from small sky patches ($f_{\rm sky}=0.003$), and that, as discussed in the main text, the posterior widths for an SO-like survey ($f_{\rm sky}\sim 0.4$) would be roughly 16 times smaller than those shown here.
    }
    \label{fig:results-noisy}
\end{figure*}

We now consider the more realistic case in which we include SO noise after component separation, as described in Sec.~\ref{sec:sims}. The resulting constraints are shown in Fig.~\ref{fig:results-noisy}. As expected, the inclusion of noise generally broadens the posterior distributions compared to the noiseless case, reducing the constraining power of all methods.

Interestingly, we find that the combination of all three MFs still provides constraints comparable to those obtained with the CNN, particularly for $\Omega_c$ and $\sigma_8$. However, the CNN now shows a more noticeable improvement over the MFs in its constraints on $\beta$, as can be seen in both the one-dimensional marginalized posterior and the joint posterior distributions involving this parameter. The power spectrum, on the other hand, continues to provide substantially weaker constraints, particularly on $\Omega_c$ and $\beta$. As in the noiseless case, $P_0$ remains poorly constrained by all methods, with broad, nearly flat marginalized posteriors.

We also investigate whether the different degeneracy directions observed in the power spectrum can be exploited to improve the constraints obtained from the MFs. To do so, we combine the three MFs with the power spectrum and train a joint posterior estimator using the framework presented in Sec.~\ref{subsec:power-spectrum}. Interestingly, this combination leads to a significant improvement, particularly in the joint posterior of cosmological parameters $\sigma_8$ and $\Omega_c$ where now they show equal constraining power to that of the CNN.

Overall, we find that the combination of all three MFs and the power spectrum recovers essentially all the cosmological constraining power of the field-level CNN, even in the presence of realistic SO noise and uncertainties in the astrophysical modeling. This suggests that these relatively simple and interpretable summary statistics, when combined, can effectively saturate the cosmological information accessible to our field-level approach, even in the presence of astrophysical uncertainties. Our findings therefore motivate further efforts towards developing robust analyses combining the power spectrum and MFs.

\section{Discussion} \label{sec:discussion}

In this work, we have investigated the total information content of the tSZ field using a CNN, and compared its constraining power with that of more interpretable summary statistics using SBI. In particular, we compared the information extracted by the CNN with that obtained from MFs and the angular power spectrum, individually and in combination. This analysis was performed on simulated tSZ maps, allowing for variations in the cluster pressure profiles as an approximate treatment of astrophysical uncertainties, and considering both noiseless maps and maps including realistic SO noise after component separation.

We find that, in the noiseless case, the CNN is able to yield tighter constraints than the summary statistics, particularly on $\Omega_c$ and $\beta$. However, when including SO-like noise, we find that the combination of MFs and the power spectrum recovers essentially all the information that the CNN is able to extract. This suggests that, at the noise levels expected for the next generation of CMB experiments, these relatively simple summary statistics could recover nearly all the information available on the field. A useful next step would be to identify which scales and morphological features drive the CNN constraints, following related interpretability studies of simulated matter fields \cite{arnab}. Nevertheless, these results provide strong motivation to explore the use of MFs, particularly in combination with the power spectrum, as a promising approach for extracting cosmological information from the tSZ field.

These results also motivate several interesting directions for future work, involving the application of the methods presented here to multi-probe $N$-body simulations such as \cite{2010ApJ...709..920S, Stein:2020its, Omori:2022uox, HalfDome}. First, our simulations are Poisson-sampled halos which include no clustering. Even though this effect is expected to be small \cite{Komatsu_1999}, an interesting next step would be to repeat this analysis using $N$-body simulations, to study the impact of the clustering terms on the constraining power of all the different methods studied here.

An important limitation of tSZ analyses on real data is foreground contamination, particularly from the CIB, which is strongly correlated with the tSZ signal and has been shown to affect its measurement significantly~\cite{PhysRevD.105.123525,Planck_tSZ_CIB,Raghunathan_2026,10.1093/mnras/staf709,tSZ_CIB_Charmetant}. Although our SO-like noise model includes residual foreground contamination after component separation, these residuals are modeled as a Gaussian random field uncorrelated with the tSZ signal. Therefore, our analysis does not capture their non-Gaussian properties or correlations with the tSZ field, whose impact on MFs remains largely unexplored. An interesting direction for future work is to investigate how the conclusions of the work presented here might change in the presence of foreground contamination, and whether the combination of higher-order statistics with power spectrum analyses could help distinguish foreground contributions from the tSZ signal.

Another caveat to our conclusions is that, even when using more realistic $N$-body simulations, tSZ maps are often generated by painting spherical pressure profiles onto halos, as we do here. However, real halos are not perfectly spherical, and different astrophysical processes can produce more complex gas distributions. These effects could be particularly relevant for morphological statistics such as $V_1$, which is sensitive to the boundaries of regions above a given threshold. Consequently, training an inference framework on simulations with spherical profiles and applying it to more complex observations could introduce biases in the inferred parameters. An important direction for future work would therefore be to test the robustness of our results using full hydrodynamical simulations (e.g., Refs.~\cite{10.1093/mnras/stac3620,10.1093/mnras/stad2419}) which naturally capture departures from spherical symmetry. Training on simulations with different implementations of baryonic physics would also allow us to assess the sensitivity of our conclusions to the details of astrophysical modeling.

Finally, while our analysis focuses on $\sigma_8$ and $\Omega_c$, the tSZ field is also sensitive to the dark energy equation of state \cite{10.1093/mnras/sty823}. Given recent hints of dynamical dark energy from DESI \cite{tr6y-kpc6,Adame_2025}, tSZ maps could provide an important independent test through the growth of cosmic structure. However, the optimal combination of summary statistics for this purpose may differ from that found here, motivating a similar analysis with simulations varying the dark energy equation of state.

All in all, we have performed field-level inference on the tSZ field and used it as a benchmark to estimate its total information content and assess how much of this information can be recovered by different summary statistics. We have found a promising combination of MFs and the angular power spectrum that recovers all the information accessible to the field-level approach in our most realistic scenario. Our results therefore provide new insights into the information content of the tSZ field and identify a promising approach (the combination of MFs with the power spectrum) for extracting this information in future cosmological analyses.

\section*{Acknowledgments}

We thank Licong Xu, David Spergel, Carolina Cuesta Lazaro, and Sebastian Wagner-Carena for helpful discussions throughout the project. 

CEV acknowledges support from the Wolfson College-CDT Studentship for Women in the Physical Sciences, and the Cambridge International Trust. 
AEB is supported by the National Science Foundation under Cooperative Agreement PHY-2019786 (The NSF AI Institute for Artificial Intelligence and Fundamental Interactions). WRC is supported by the Science and Technology Facilities Council (STFC) through an Ernest Rutherford Fellowship (UKRI2424: “Astrophysics and Cosmology from CMB Scattering”) and through the Royal Society grant ICA\textbackslash R2\textbackslash 252140. 
AC acknowledges support from STFC (grant references ST/X006387/1 and UKRI1164).
JCH and AS acknowledge support from NASA grant 80NSSC23K0463 [ADAP].
NB and JCH acknowledge support from NASA grant 80NSSC22K0721 [ATP].
Model training and inference were performed using facilities supported by the Scientific Computing Core at the Flatiron Institute, a division of the Simons Foundation.
The authors acknowledge the Texas Advanced Computing Center (TACC) at The University of Texas at Austin for providing computational resources that have contributed to the research results reported within this paper. This research used resources of the National Energy Research Scientific Computing Center (NERSC), a Department of Energy User Facility.  This is not an official Simons Observatory collaboration paper.

\bibliography{main}{}
\bibliographystyle{apsrev4-2}

%

\vspace{5mm}




\appendix

\section{Fixed astrophysics}\label{app:results-cosmo}

In this appendix, we present results for the idealized scenario in which the astrophysical parameters governing the cluster pressure profiles are held fixed, and only the cosmological parameters $\Omega_c$ and $\sigma_8$ vary. These simulations are described in Sec.~\ref{sec:sims}, and have narrower prior ranges than those used in our main analysis.  We compare the constraints obtained from field-level CNN inference with those from the three MFs, and their combination. In this case, we compute the MFs using 100 linearly spaced thresholds over the range $-1\leq\nu\leq3$. We include realistic SO-like noise in our training simulations as described in Sec.~\ref{sec:sims}. 

\begin{figure}
    \centering 
    \includegraphics[]{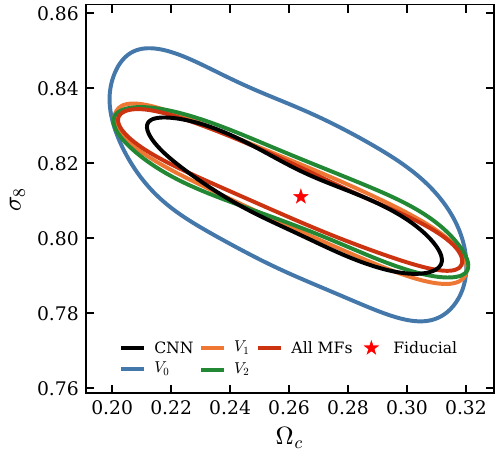}
    \caption{68\% contours of the posterior distributions for $\Omega_c$ and $\sigma_8$, obtained from the CNN (black), the three Minkowski functionals individually ($V_0$, $V_1$, and $V_2$), and their combination (red). The posteriors are trained on simulations including realistic SO-like noise after component separation, with the astrophysical parameters held fixed, and then evaluated on 1,000 independent tSZ maps with the same fiducial cosmology: $(\Omega_c,\sigma_8)=(0.264,0.811)$. The equally weighted mixture of the 1,000 posterior distributions are shown.
    }
    \label{fig:results-cosmo-noisy}
\end{figure}

The resulting constraints are shown in Fig.~\ref{fig:results-cosmo-noisy}. Interestingly, the combination of all three MFs yields constraints very similar to those obtained with the CNN. Individually, $V_1$ and $V_2$ already capture much of the constraining power of the combined MFs. In contrast, $V_0$ produces considerably weaker constraints. In all cases, we find a strong degeneracy between $\sigma_8$ and $\Omega_c$. Both the CNN and the MFs recover very similar degeneracy directions, suggesting that the MFs are already sensitive to the underlying combination of cosmological parameters to which the tSZ field is intrinsically most sensitive.

A similar analysis using the same simulation framework was carried out in Ref.~\cite{Alina}, which investigated the constraining power of MFs and other non-Gaussian summary statistics. Unlike our analysis, that work employed a Gaussian likelihood, rather than a normalizing flow, and considered map patches with slightly different angular resolution. Despite these differences, we recover a similar degeneracy direction and comparable constraining power in the joint $(\sigma_8,\Omega_c)$ posterior.

\section{Coverage tests} \label{app:coverage-tests}

To check the reliability of our inferred posterior distributions, we perform coverage tests using Tests of Accuracy with Random Points (TARP) \cite{pmlr-v202-lemos23a}. Coverage tests assess whether the inferred posterior distributions are well calibrated, or whether they underestimate or overestimate the uncertainties on the parameters. For a well-calibrated posterior, a region containing 68\% of the posterior probability should enclose the true parameters in approximately 68\% of the independent realizations used in the test. Here, we perform these using the 1205 simulations in the validation set.

We use TARP to test the calibration of the full four-dimensional posterior over $(\Omega_c,\sigma_8,P_0,\beta)$, rather than testing each parameter individually. For each independent validation simulation, we draw samples from the inferred posterior and select a random reference point in the parameter space. We then calculate the fraction of posterior samples that lie closer to this reference point than the true parameters. Distances are computed in the four-dimensional parameter space, with each parameter normalized by the width of its sampled range. For a well-calibrated posterior, the true parameters should behave statistically like another draw from the posterior distribution. As a result, plotting the empirical coverage against the nominal credibility should give a diagonal line, showing that the inferred credible regions contain the true parameters with the expected frequency.

\subsection{Posterior calibration across methods}

\begin{figure*}
    \centering 
    \includegraphics[width=0.8\linewidth]{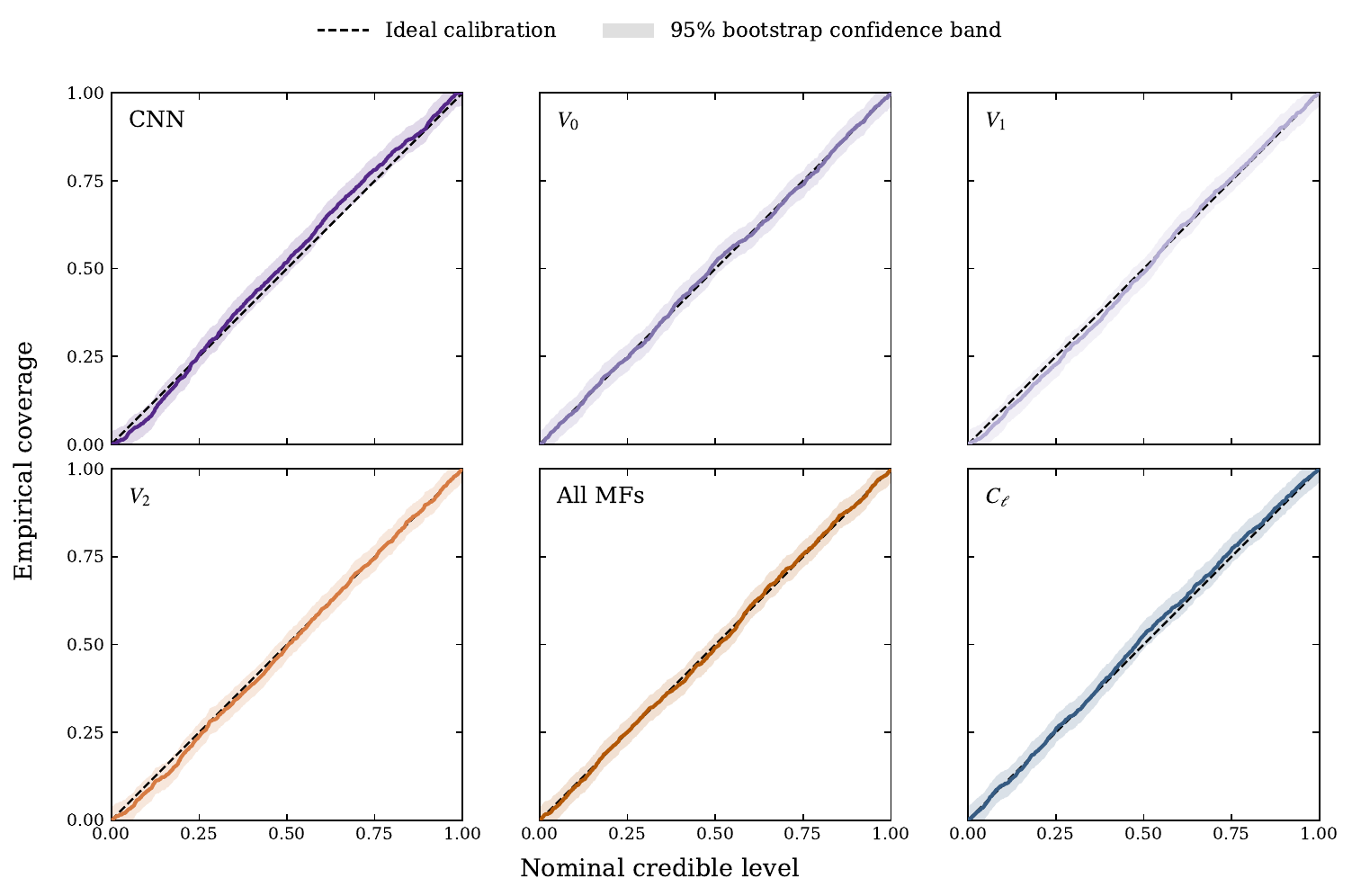}
    \caption{TARP coverage tests for the six inference methods considered in the noiseless case. All show good posterior calibration, with coverage curves following the ideal diagonal within the 95\% bootstrap confidence bands.}
    \label{fig:tarp-noiseless}
\end{figure*}

Here, we present the TARP coverage tests for the six inference methods presented in the main text for the noiseless case, whose posteriors are shown in Fig.~\ref{fig:results-noiseless}. As can be seen in Fig.~\ref{fig:tarp-noiseless}, all methods exhibit good calibration, with the coverage curves following the ideal diagonal and remaining within the expected 95\% confidence band obtained using bootstrap resampling.

\subsection{Dependence of posterior calibration on training-set size}

\begin{figure*}
    \centering 
    \includegraphics[width=0.8\linewidth]{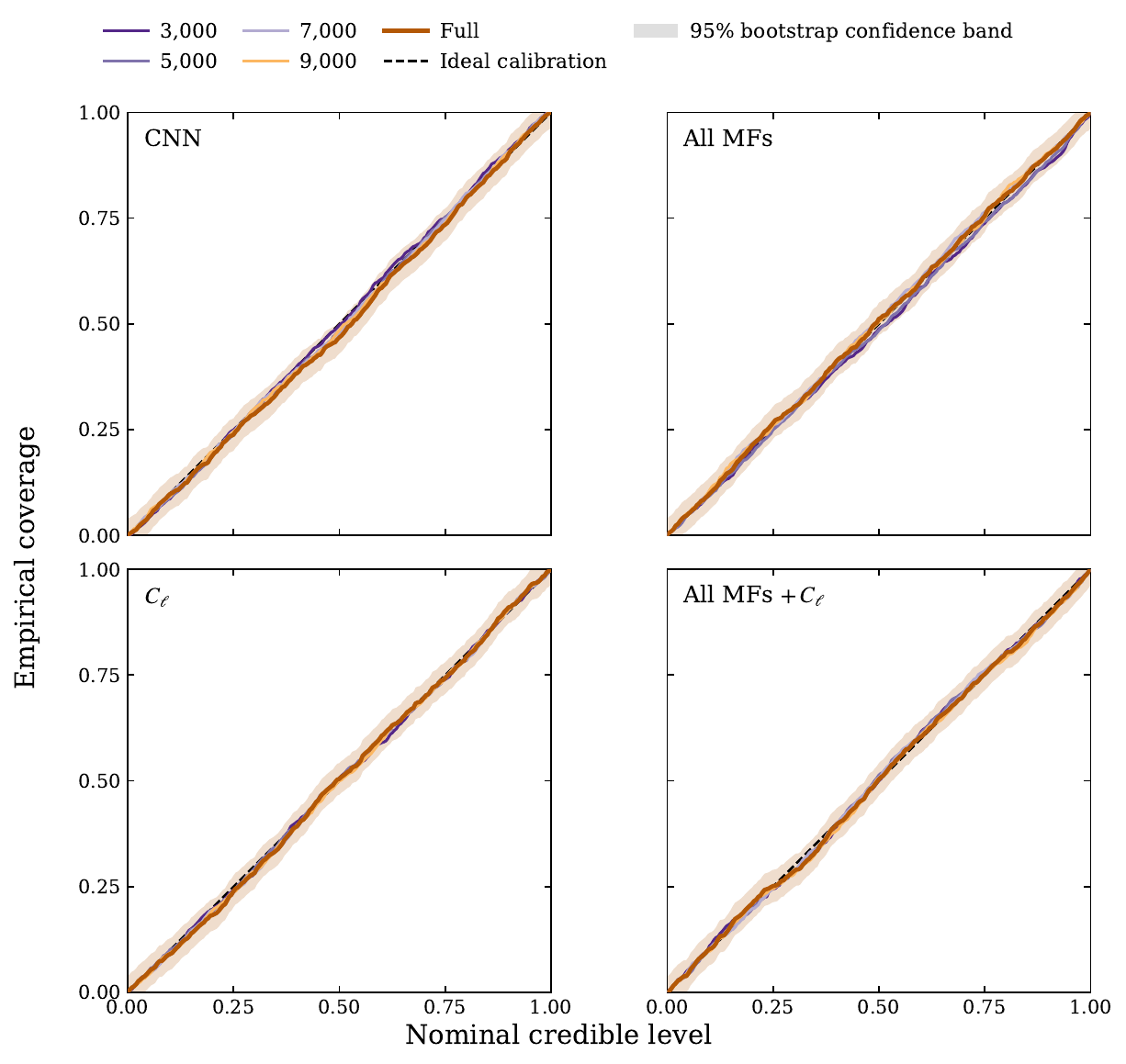}
    \caption{TARP coverage tests for the four inference methods considered in the main text for the noisy case, using different numbers of training simulations. All methods show good posterior calibration across training-set sizes, with coverage curves consistent with the diagonal within the 95\% bootstrap confidence bands.}
    \label{fig:tarp-noisy}
\end{figure*}

We also investigate whether the number of training simulations is sufficient to obtain reliable posterior distributions. To do this, we repeat the TARP coverage tests using different training-set sizes, ranging from 3,000 simulations to the full training set. We focus on the four inference methods considered in the noisy case in the main text: the CNN, all MFs, the power spectrum, and their combination. We show these for the case with noise in Fig.~\ref{fig:tarp-noisy}, where all methods exhibit good calibration across the different training-set sizes, with coverage curves following the diagonal and showing no significant deviations within the 95\% bootstrap confidence bands. We note that we repeat this check for the noiseless case and find the same results. This shows that our training sets are sufficiently large to achieve well-calibrated posterior distributions, and that no additional simulations are needed to improve calibration.

\end{document}